# Surface Bubble Growth in Plasmonic Nanoparticle Suspension

Qiushi Zhang[1], Robert Douglas Neal[2], Dezhao Huang[1], Svetlana Neretina[2,3], Eungkyu Lee[1*], and Tengfei Luo[1,3,4*]

[1]Department of Aerospace and Mechanical Engineering, University of Notre Dame, Notre Dame, IN, USA

[2]College of Engineering, University of Notre Dame, Notre Dame, Indiana, United States

[3]Department of Chemical and Biomolecular Engineering, University of Notre Dame, Notre Dame, USA.

[4]Center for Sustainable Energy of Notre Dame (ND Energy), University of Notre Dame, Notre Dame, USA.

*Correspondence to: tluo@nd.edu; elee18@nd.edu

## Abstract

Understanding the growth dynamics of the microbubbles produced by plasmonic heating can benefit a wide range of applications like microfluidics, catalysis, micro-patterning and photo-thermal energy conversion. Usually, surface plasmonic bubbles are generated on plasmonic structures pre-deposited on the surface subject to laser heating. In this work, we investigate the growth dynamics of surface microbubbles generated in plasmonic nanoparticle (NP) suspension. We observe much faster bubble growth rates compared to those in pure water with surface plasmonic structures. Our analyses show that the volumetric heating effect around the surface bubble due to the existence of NPs in the suspension is the key to explain this difference. Such volumetric heating increases the temperature around the surface bubble more efficiently compared to surface heating which enhances the expelling of dissolved gas. We also find that the bubble growth rates can be tuned in a very wide range by changing the concentration of NPs, besides laser power and dissolved gas concentration.

**Introduction**

Plasmonic bubbles can be generated in noble metal plasmonic nanoparticles (NPs) suspensions upon the irradiation of pulsed laser due to the enhanced plasmonic resonance [1-6]. These micro-sized bubbles can play important roles in a wide range of applications, including biomedical imaging [7–10], healthcare diagnosis [11-15], and microfluidic bubble logics [16]. In recent years, studies on the growth dynamics of plasmonic surface bubbles have attracted significant attention [6, 17-20]. As discussed before [17], the growth of surface bubbles can be generally divided into two phases, i.e., short-time and long-time growth phases. In the short-time growth phase (phase I), the surface bubble experiences an explosive nucleation due to the vaporization of the liquid surrounding NPs on the surface. In the long-time growth phase (phase II), the bubble growth is mainly because of the expelling of dissolved gas from the liquid surrounding the nucleated surface bubbles.

To study the growth dynamics of plasmonic surface bubbles, substrates with pre-decorated plasmonic metal nanoclusters submerged in de-ionized (DI) water have been commonly used [6, 17-22]. In this type of experimental systems, surface bubbles usually have an extremely short (10 to hundreds of milliseconds) phase I, in which the bubbles grows very fast ($10^6 \sim 10^7$ μm$^3$/s), and the volume growth is proportional to $\sqrt{t}$, where $t$ is time. Compared to phase I, phase II lasts much longer (e.g., bubble can be stable for minutes or even hours depending on the growth environment), and the volume growth of surface bubble is much slower and linear in time ($\sim 10^4$ μm$^3$/s). Although the two growth phases are different, the basic cause is the same – plasmonic heating. In addition to Ref. [17], different behaviors and mechanisms of plasmonic surface bubbles have been investigated and proposed in other studies. For example, Wang et al. have revealed the giant and oscillating plasmonic surface bubble in the very early life phase. This is due to the composition of the surface bubble in the early life phase gradually changing from vapor to gas [18]. Baffou et al. have studied the bubble shrinkage behavior. Since the surface bubble in the long-time growth phase is mainly made of dissolved gas, it displays a linear and slow volume shrinkage [19]. Liu et al. [6] and Chen et al. [20] have studied the bubble growth dynamics on nano-arrays. They find that the volume growth rate of bubble is largely related to the density and geometry of these nano-arrays, which determine the collective input heating power. Zhao et al. [23] and Lin et al. [24] have shown the convective flow around surface bubble can be used to trap nano- or microparticles. This originates from the Marangoni flow surrounding the surface bubble. This Marangoni flow is attributed to the temperature gradient formed around the micro-size surface bubble [21, 25].

In addition to pre-deposited nanostructures, generating surface bubble directly using the plasmonic heating of NP suspension has also been demonstrated [26-28]. In Richardson et al.'s work [26], the theoretical model of light-to-heat conversion efficiency in NP suspension is established from fitting the experimental data of a droplet in the millimeter-scale. The adsorption and conversion efficiencies highly depend on the concentration of NPs and input laser power. Armon et al. [27] have demonstrated that bubble movement in NP suspension can be used for micro-patterning. Compared to pre-deposited optically resistive nanostructures, plasmonic NP suspensions features the advantages of simpler processes, higher heating efficiency and potentially better compatibility with biological environments. Fundamentally, plasmonic NP suspension is subjected to volumetric heating wherever the excitation laser beam covers instead of only surface heating as in the pre-deposited nanostructure cases. However, the detailed investigation of the plasmonic surface bubble growth in NP suspensions have not been performed, which will be important for its potential applications.

In this paper, we systematically study the growth dynamics of surface bubbles in plasmonic NP suspensions via experiments accompanied with theoretical analyses. Micro-sized plasmonic surface bubbles are generated with both pre-deposited NPs clusters and NP suspensions under the irradiation of pulsed laser at the surface plasmon resonance (SPR) peak of the NPs. The growth dynamics of the surface bubbles in both conditions are investigated and compared using high-speed videography. It is demonstrated that under the same laser conditions, the surface bubbles grow much faster in the NP suspensions than in DI water with pre-deposited nanoparticles. Our analysis indicates that it is the volumetric heating in the NP suspension that leads to a higher heating efficiency, which results in higher temperature around the surface bubble and thus larger bubble growth rates. In addition, we also find that the bubble growth rates can be tuned by changing the concentration of NPs very efficiently, besides laser power and dissolved gas concentration.

**Results and discussion**

We first study the plasmonic surface bubble growth dynamics in two comparing cases. In Case I, we generate micro-sized surface bubbles on a bare quartz surface immersed in a NP suspension, as shown in Figure 1a. In Case II, the bubbles are generated on a quartz surface pre-deposited with NP clusters immersed in DI water, as shown in Figure 1d. In both cases, pulsed laser excitations are used, and the beams are focused on the inner surface of the quartz substrate (see Experimental Methods section for details). In case I, a surface bubble nucleates in a few seconds upon laser irradiation. During the short period before bubble nucleation, a small amount of NPs are found deposited on the quartz surface as shown in the SEM image in Figure 1b (also see Supplementary Information, Figure S1). The NPs are deposited due to optical forces as recently revealed in Ref. [29] and can work as hot spots and nucleation centers for the surface bubble generation. We notice that the area with deposited NPs on the quartz surface is about ~ 100 $\mu m^2$, comparable to the laser beam cross-sectional area. When using the 10x objective lens, the diameter of our Gaussian laser spot is ~ 11 μm as determined from a beam profiler. This means that once the surface bubble nucleates and grows, these deposited NPs will be mostly in contact with the gaseous phase, which limits their effectiveness of heating up the liquid in the phase II growth due to the large thermal resistance of the gaseous phase [30-32]. As described in Figure 1a, besides the deposited NPs serving as a surface heater, there is another heating source in the NP suspension that is the volumetric heating of the laser beam covered area due to absorption from the suspended NPs. These NPs can provide additional heat to the liquid around the surface bubble during the whole growth period.

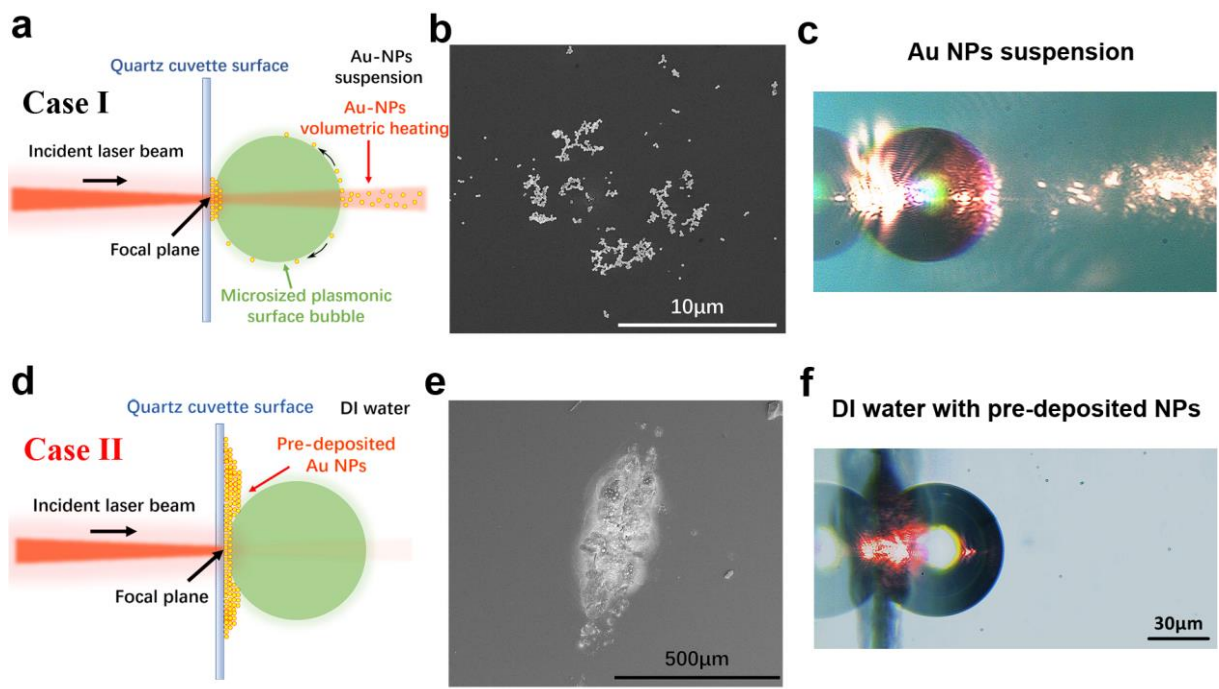

**Figure 1.** Schematic descriptions of a micro-sized plasmonic surface bubble grows in (**a**) Au NP suspension (Case I) and (**d**) DI water with pre-deposited NPs on surface (Case II). Scanning electron microscope (SEM) images of pre-deposited Au NPs at the bubble nucleation site in (**b**) Case I and (**e**) Case II. Optical images from the side view of a plasmonic surface bubble under laser illumination in (**c**) Case I and (**f**) Case II. Scale bar is the same in (**c**) and (**f**). The bright regions in (**c**) and (**f**) are from the laser scattered by either pre-deposited or suspended Au NPs.

On the other hand, the condition where surface bubble grows in Case II (DI water with pre-deposited NPs on surface) has two major differences compared with Case I, as illustrated in Figure 1d. Firstly, Case II has a much larger amount of NPs pre-deposited on the surface, which can lead to stronger surface heating. As shown in Figure 1e (also in Figure S1), it is easy to see there are much more Au NPs pre-deposited on the surface in Case II than in Case I. Secondly, since the surface bubble is surrounded by DI water rather than NP suspension in Case II, there is no volumetric heating, leaving surface heating as the only heating source. This can be visually observed from the glowing spots in the optical images of surface bubbles under laser illumination, as shown in Figures 1c and f. These glowing spots correspond to the scattered light from the plasmonic Au NPs, either deposited on the surface or suspended in liquid. As seen from Figures 1c and f, there are glowing spots both on the surface and in the laser beam covered volume on top of the bubble in Case I, while there are only such spots on the surface in Case II.

Since the two cases have distinct heating geometries, different bubble growth behaviors are expected. We record and compare the bubble growth dynamics in the two cases using high-speed videography when they are subject to the same laser irradiation conditions. Recall that surface bubble growth experiences two phases, i.e., the explosive vaporization (phase I) and gas expelling (phase II). As shown in Figure 2a, the bubble in Case II undergoes a very fast growth in phase I, with the duration of shorter than 500 ms. The reason of this fast growth is that the large amount of heat from the highly dense NPs pre-deposited at the surface in Case II can quickly lead to a high surface temperature to vaporize water. After the bubble contact line circle is larger than the laser spot size as the bubble grows bigger, the heated pre-deposited NPs can no longer maintain the liquid-vapor interface of the bubble above the vaporization temperature due to the large vapor thermal resistance. Then the bubble growth slows down and transition into phase II, which is displayed as a kink in the volume growth plot (Figures 2a and b). On the other hand, in the NP suspension (Case I), the phase I bubble grows much slower than that in Case II, which can be attributed to the much fewer NPs on the surface as heating sources (see Figures 1b and e). However, it is interesting to see that in the NP suspension, the bubble has longer phase I growth (~ 3s) and reaches a larger size at the end of this period. This is likely due to that the volumetric effect in NP suspension can provide higher heating efficiency than surface heating, which is shown in later discussions. The higher heating efficiency can maintain the evaporation of the water surrounding surface bubble at a larger bubble size. During this longer phase I, the oscillations of the bubble volume are also observed, which is similar to the behaviors in Ref. [18].

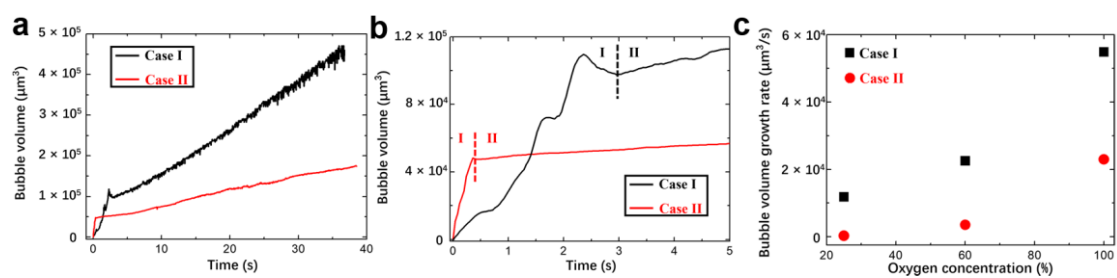

**Figure 2.** **(a)** Surface bubble volume as a function of time in the two cases. Both are from liquids with 60% degassing level and the same laser power of ~ 1.1 W. **(b)** The plot shows a zoomed view of the range from 0 to 5 s in **(a)**. **(c)** The averaged surface bubble volume growth rates of phase II in the two cases under different degassing levels.

Phase II growth usually lasts much longer than phase I. As shown in Figure 2a, both cases have linear volume growth in phase II, consistent with the growth behavior of phase II gas bubbles in previous works [17]. However, there is a clear difference between the growth rates in the two cases, with the NP suspension showing a much higher growth rate. Since phase II growth is due to dissolved gas expelling at elevated temperatures, we then have performed the same experiments but with different degassing levels (see Experimental Methods section for details). As shown in Figure 2c, the phase II bubbles always grow faster in the NP suspension (Case I) than in DI water with pre-deposited NPs (Case II) disregard the degassing levels. When the dissolved gas is reduced to a very low level (e.g., ~ 25%), the surface bubble in Case I still maintains a significant growth rate, while that in Case II hardly grows.

To reveal the mechanism of the different phase II growth behaviors, we have first confirmed that the compositions of the bubble in both cases are dissolved gas rather than vapor. As the bubble shrinkage study in Supplementary Information (SI2) evidenced, the shrinkage of our plasmonic surface bubble in phase II is very slow, taking more than 30 mins. If it is a vapor bubble, it would collapse immediately (in milliseconds) when the thermal excitation is turned off [33]. Moreover, our bubble shrinkage is linear with a very slow bubble shrinkage rate (~ 420 $\mu m^3/s$, see Supplementary Information, Figure S3). This shrinkage rate is of the same order of magnitude as reported in Ref. [19], and it is proved that the bubble shrinkage in our cases has a feature of expelled gas re-dissolving into liquid as the temperature around the bubble slowly decreases. With this confirmed, we examined the difference in heating sources (i.e., surface heating and volumetric heating) that influences the dissolved gas expelling rates. Since the surface heating is different in the two cases given the drastically different NP densities on the surface (see Figures 1b and 1e), we studied a third case where we immersed the substrate with pre-deposited NPs in the NP suspension (Case III) to better quantify the role of volumetric heating. With the same laser power of 1.1 W, we observed a much faster phase II bubble growth rate in Case III than in Case II (Figure 3a). By taking the difference of the phase II bubble growth rates (K) of these two cases, the volume growth rate that can be attributed to volumetric heating in the suspension is ~ $4\times10^4$ μm³/s. This is more than two times larger than the growth rate by solely surface heating. For a phase II bubble, the mass influx of dissolved gas into the bubble ($dm_g$) is proportional to the change in local oversaturation ($d\zeta$) by the following formula [17]:

$$dm_g = C_s V_w d\zeta \tag{1}$$

where $C_s$ is the local air solubility in water, and $V_w$ is the volume of water contributing to the gas expelling for bubble growth, which depends on the thermal boundary layer thickness [34] at the bubble surface. $d\zeta$ is further proportional to the change in the local temperature surrounding the bubble ($dT$) by:

$$d\zeta = -\frac{C_\infty}{C_s^2}\frac{dC_s}{dT}dT \tag{2}$$

where $C_\infty$ is the gas saturation far away from the bubble. Combining equations (1) and (2), it is clear that the increase in the temperature of liquid water surrounding the surface bubble (boundary layer) [34] is the main cause of the phase II bubble growth.

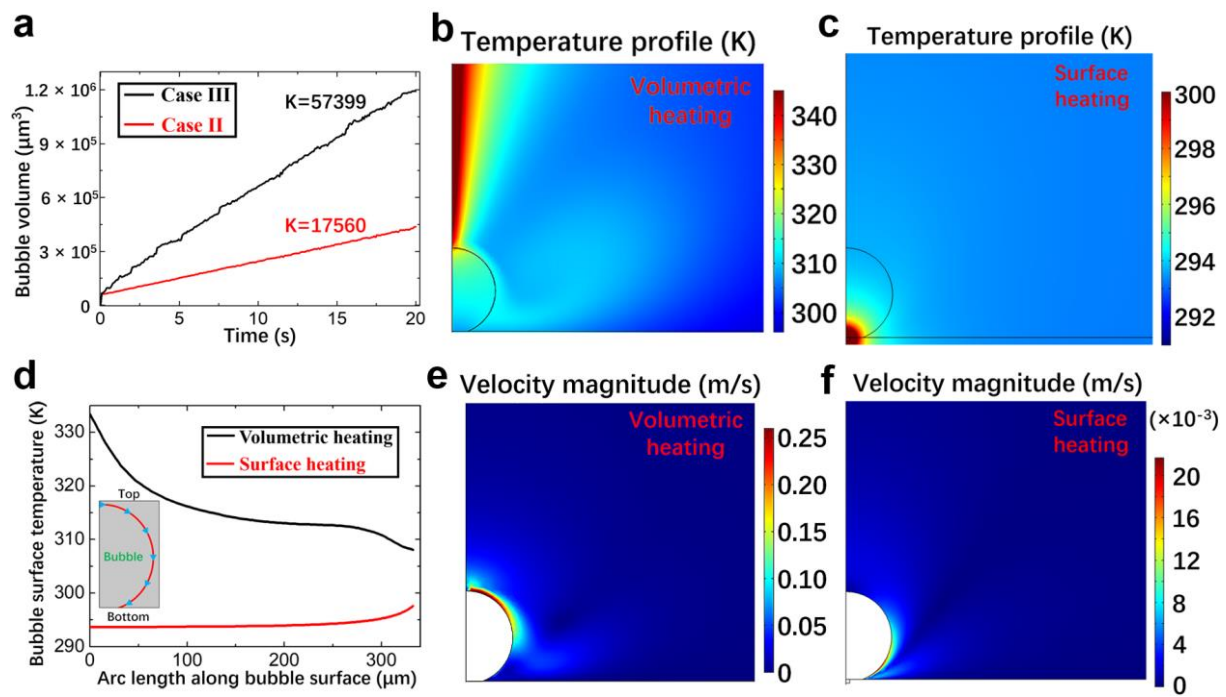

**Figure 3. (a)** Surface bubble volume as a function of time in DI water (Case II) and NP suspension (Case III), both with the same amount of pre-deposited Au NPs and the same laser power of ~ 1.1 W. The volume growth rates (K) in phase II are shown in the plots. The simulated temperature profiles in the **(b)** volumetric heating and **(c)** surface heating cases. **(d)** The simulated bubble surface temperature from the top to the bottom of the bubble in the two cases. The simulated liquid flow velocity magnitude contours in the **(e)** volumetric heating and **(f)** surface heating cases.

To quantify the volumetric heating effect on the temperature around the bubble, we employ finite element simulations to investigate the temperature distribution under the two different heating geometries (see simulation details in the Supplementary Information, SI3). The simulated temperature profiles of the two heating conditions are shown in Figures 3b and c. We can easily observe the difference in the locations and distributions of the hottest regions in the two cases as they are around the respective heating sources. The temperatures around the bubble surfaces are also different. Figure 3d shows the temperature at the bubble surfaces as a function of the arc length from the top to the bottom of the bubble. The overall bubble surface temperature in the volumetric heating case is higher than the one in the surface heating case, with the average temperature of the former 20 K higher than the latter. The reasons of this surface temperature difference are as follow: 1. In the surface heating case, there is significant heat loss from the heating source to the quartz substrate; 2. Most of the surface heater is in contact with the gas in the bubble, so the heat cannot be conducted to the bubble surface efficiently; 3. In the volumetric heating case, the thermocapillary flow of liquid near the surface of the bubble help distribute heat around the bubble surface (see the velocity profile in Figures 3e and f). These simulation results indicate that volumetric heating is much more efficient in heating the surrounding of the bubble to a higher surface temperature, and this should be the main cause of the dramatically increased bubble growth rate.

The volumetric heating in the NP suspension provides additional means to control surface bubble growth via tuning the NP concentration in suspension, besides the conventional laser power control in surface heating methods. To study the effects of changing laser power, we tune the source laser power from 0.3 W to 1.12 W, which starts from the minimum laser power that can enable bubble nucleation to the maximum power achievable in our laser system. As plotted in Figures 4a and b, the volume growth rate has only been increased by less than two times in this laser power range which changed four times. To examine the effects of changing Au NP concentration, we prepare the Au NP suspensions with four different NP concentrations from ~ $1\times10^{15}$ to $4\times10^{15}$ particles/m$^3$ (see Experimental details in the Experimental Methods section). Other experimental conditions, like laser power (~ 1.1 W) and dissolved air concentration (100%), are kept the same in all experiments. The phase II bubble volume growth as a function of time for the four different NP concentrations are all linear, but with significantly different slopes, and the growth rates are shown in Figure 4d. As can be seen, the bubble volume growth rate is highly sensitive to the change of Au NP concentration, increasing by one order of magnitude with a 4-fold increase in NP concentration. This is a much more significant change compared to the effect from tuning the laser power. This will be beneficial for many microfluidics applications which desire widely tunable bubble sizes [16].

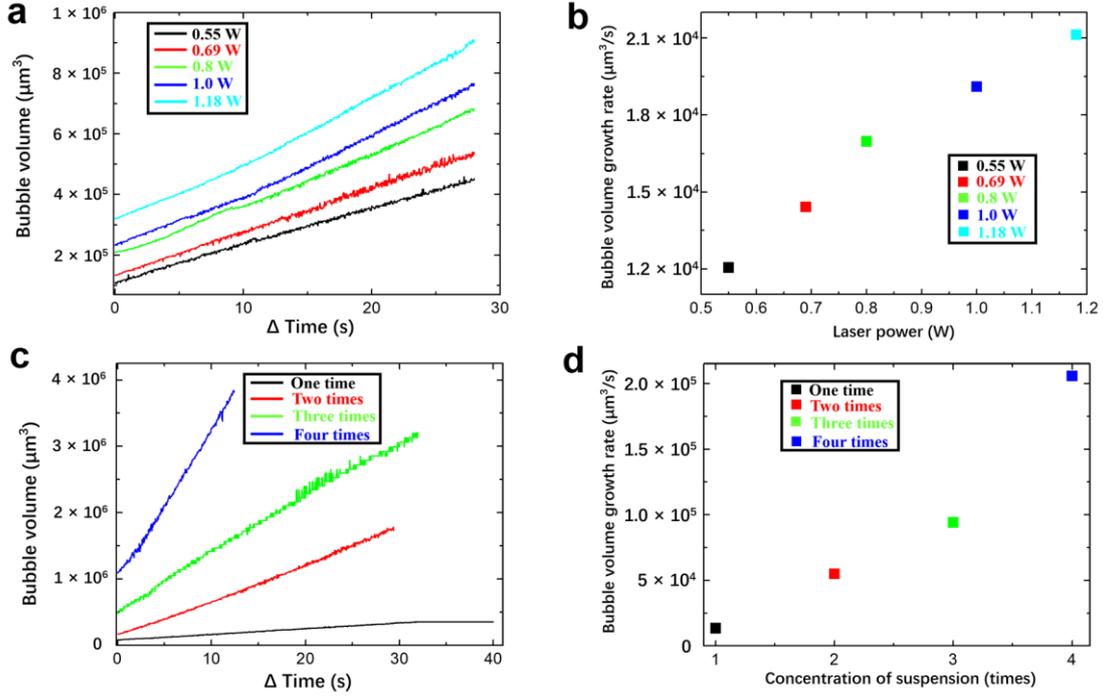

**Figure 4.** (**a**) Phase II surface bubble volume growth as a function of time under different laser powers from 0.3 W to 1.12W. The dissolved air concentration is 100%, and the Au NP concentration is $1 \times 10^{15}$ particles/m$^3$ for all cases. (**b**) The averaged phase II surface bubble volume growth rates under different laser powers. (**c**) Phase II surface bubble volume growth as a function of time in the NP suspensions with different concentrations of Au NPs. The dissolved air concentration of 100% and laser power of ~ 1.1 W are held constant for all cases. (**d**) The averaged phase II surface bubble volume growth rates in the NP suspensions with different concentrations of Au NPs.

**Conclusion**

To summarize, the growth dynamics of plasmonic surface bubbles in the two cases, NP suspension (Case I) and DI water with pre-deposited NPs on surface (Case II), have been systematically investigated in this work. Due to the special volumetric heating geometry, NP suspension enables much higher bubble volume growth rates compared to the more conventional surface heating conditions. This is mainly because that the volumetric heating geometry has higher heating efficiency and is able to maintain a higher bubble surface temperature under the same laser power. We have also demonstrated that NP suspension can provide greater bubble growth tunability via changing the NP concentration. These results may provide fundamental insights to surface bubble growth dynamics in plasmonic suspensions. They may also offer additional degrees of freedom to control surface bubbles for microfluidics applications.

**Experimental Methods**

**Pulsed laser:** The mode-locked monochromatic femtosecond pulsed laser we used in our experiments is emitted from a Ti:Sapphire crystal in an optical cavity (Spectra Physics, Tsunami). The laser has a center wavelength of 800.32 nm and a full-width-half-maximum length of ~ 10.5 nm. The laser power is in the range of 0.3 ~ 1.2 W with the pulse duration of ~ 200 fs and the repetition rate of 80.7 MHz. The laser beam is guided by a series of broadband dielectric mirrors and finally focused by a 10× (Edmund Optics) objective lens to achieve a Gaussian intensity profile with a $1/e^2$ radius of 20 μm on the quartz surface. An optical shutter controlled by a digital controller (KDC101, Thorlabs) is used to turn on/off the laser (see Figure 5a).

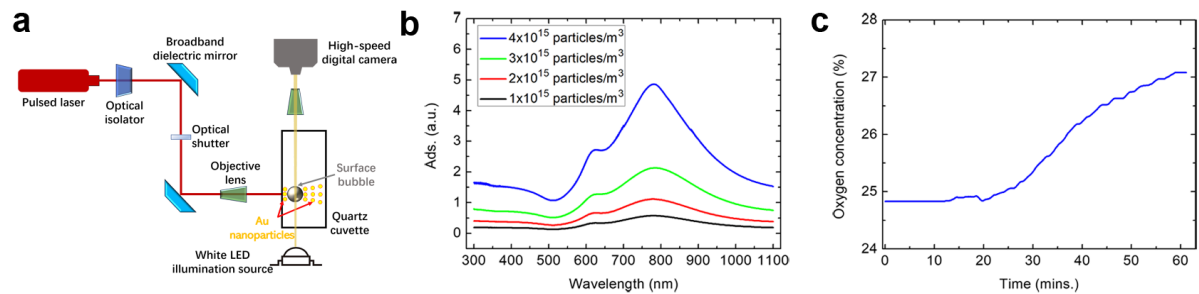

**Figure 5.** (**a**) Schematic of the experimental setup to characterize the growth dynamics of surface bubbles. (**b**) Measured absorption spectra for determining the Au NP concentrations in each suspension. The adsorption spectra of the suspensions with 1 to $4 \times 10^{15}$ particles/m$^3$ Au NP concentrations are plotted. (**c**) Oxygen concentration as a function of time of a degassed suspension as measured by an oxygen sensor. The degassing level can be kept in normal pressure for more than an hour with less than 5% increase.

**High-speed videography:** The plasmonic surface bubble growth dynamics is recorded using a digital camera (HX-7, NAC). The digital camera is aligned to record the surface bubbles from the side view. A white LED illumination source and a 20× objective lens (Edmund Optics) are used in the video recording process (see Figure 5a). The videos recorded by the digital camera are then analyzed in a home-built MATLAB code, where the size of the surface bubble at each time frame is fitted. With these fitted results, we can then plot bubble volume as a function of time.

**Au NP suspension preparation:** The plasmonic Au NP suspension is prepared by ultrasonic dispersion of spherical Au NPs (Nanospectra Bioscience, Inc) consisting of a silica core (~ 50 nm of radius) and an Au shell (~ 10 nm of thickness) in DI water. The resonant wavelength of the Au NPs in suspension is around 780 ~ 800 nm (average peak 785 nm) (see Figure 5b), and the concentration varies from case to case. In our experiment, the plasmonic NP suspension is contained in a quartz cuvette (Hellma, Sigma-Aldrich, 10 mm × 10 mm). Before filled with suspension, the quartz cuvette is cleaned in an ultrasonic bath and dried at 150 °C for 10 min.

**NP pre-deposition:** To pre-deposit a significant amount of Au NPs on the quartz surface, we first used pulsed laser to generate a large surface bubble (radius ~ 200 μm) on the surface in contact with the plasmonic NP suspension. Then, we generate another smaller surface bubble (radius ~ 30 μm) very close to the large bubble. The growth of the smaller bubble is restricted by the nearby large one, as it will be swallow by the larger bubble when they contact each other (See Movie S1). After that, a small amount of NPs can be deposited on the surface and another smaller bubble will be generated at the same site shortly. By repeating this process tens of or even over a hundred times, we can eventually deposit a large amount of NPs on the quartz surface (see Figure 1e). Once we have accumulated significant amount of deposited NPs, we replaced the plasmonic NP suspension in the cuvette with DI water to prepare the experiment for Case II. We note the specific technique to deposit the large amount of the NPs does influence the validity of the conclusion in the paper.

**Water degassing:** The Au NP suspension is degassed in a sealed chamber pumped by an external mechanical pump. The concentration of oxygen in the suspension is measured by an oxygen sensor, which is used to quantify the concentration of dissolved air. The concentration of oxygen is ~ 8.3 mg/L in the suspension without degassing. After 3h degassing, the concentration of oxygen becomes ~ 60% of the original concentration; after 24h degassing, the concentration of oxygen drops to ~ 25%. During experiments, the quartz cuvette containing degassed suspension is kept sealed to slow down the air re-dissolving process. Based on our tests, the concentration of oxygen increases less than 5% within 1.5h while kept sealed in air (See Figure 5c). Since each of our experimental measurement normally lasts for less than 15 mins, the concentration of oxygen in degassed suspension is considered to be constant.

**Acknowledgement**

This work is supported by National Science Foundation (1706039) and the Center for the Advancement of Science in Space (GA-2018-268). T.L. would also like to thank the support from the Dorini Family endowed professorship in energy studies.

# Supplementary Information for

# Surface Bubble Growth in Plasmonic Nanoparticle Suspension

Qiushi Zhang[1], Robert Douglas Neal[2], Dezhao Huang[1], Svetlana Neretina[2,3], Eungkyu Lee[1*], and Tengfei Luo[1,3,4*]

[1]Department of Aerospace and Mechanical Engineering, University of Notre Dame, Notre Dame, IN, USA

[2]College of Engineering, University of Notre Dame, Notre Dame, Indiana, United States

[3]Department of Chemical and Biomolecular Engineering, University of Notre Dame, Notre Dame, USA.

[4]Center for Sustainable Energy of Notre Dame (ND Energy), University of Notre Dame, Notre Dame, USA.

*Correspondence to: tluo@nd.edu; elee18@nd.edu

**SI1. Characterizing the deposited Au NPs on quartz surface**

Figures S1a and b show the side-view optical images of the location where surface bubble nucleates and grows in the two cases. Panels **(a)** and **(b)** share the same scale bar. It is clear that Case II has much more Au NPs pre-deposited on the quartz surface.

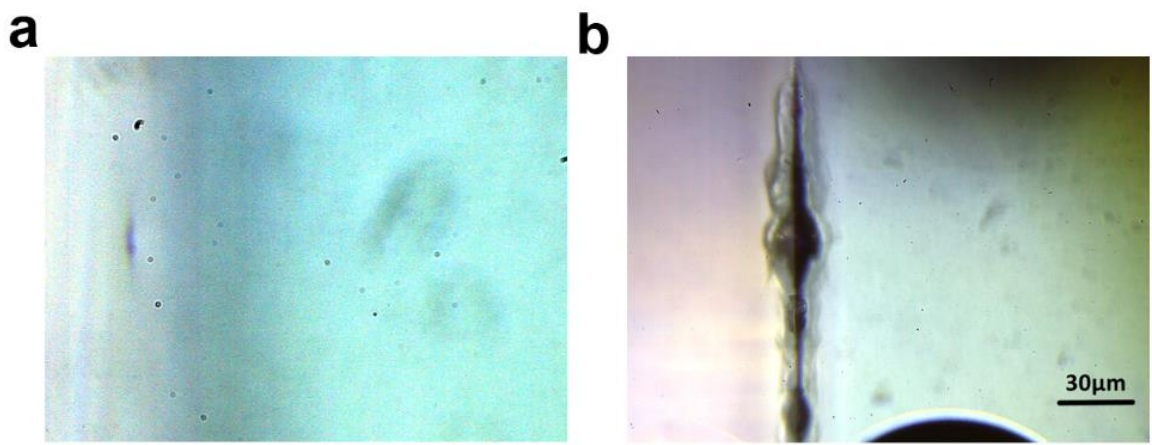

**Figure S1.** Optical images of the Au NPs deposited on the inner surface of the quartz substrate, where the laser focal plane resides, right before the surface bubble nucleation in **(a)** Case I and **(b)** Case II (scale bar: 30μm).

Figure S2 shows the scanning electron microscope (SEM) studies of the pre-deposited Au NPs at the bubble nucleation site in Case II. The sample is coated with Iridium (Ir) thin film of ~ 0.5 nm for SEM. The imaging parameters of Figure 1b (in main text) are: Current is 13 pA; Voltage is 5 kV; Magnification is 12000×; working distance is 4.2 mm. The imaging parameters of Figure 1e (in main text) are: Current is 50 pA; Voltage is 5 kV; Magnification is 250×; working distance is 4.8 mm. The imaging parameters of Figure S2a are: Current is 50 pA; Voltage is 5 kV; Magnification is 160000×; working distance is 4.1 mm. The imaging parameters of Figures S2b and d are: Current is 50 pA; Voltage is 10 kV; Magnification is 80000×; working distance is 4.3 mm. The imaging parameters of Figure S2c are: Current is 25 pA; Voltage is 5 kV; Magnification is 50000×; working distance is 4.1 mm. Energy-dispersive X-ray (EDX) spectrum takes the data from an area of 2.5×2.5 μm$^2$ with pre-deposited Au NPs (5 kV, 0.8nA). As shown from Figures S2a to b, we can observe the melting and merging of individual Au NP. Then in Figure S2c, some Au clusters are found to form. Figure S2d is the back-scattered SEM image confirming the melting and merging of the Au shells of NPs. Figure S2e shows the Energy-dispersive X-ray (EDX) spectrum of the pre-deposited Au NPs. The Au peak is from the deposited Au clusters. The Si peak is from both the silica core of Au NPs and quartz substrate. The Ir peak is from the coating for SEM.

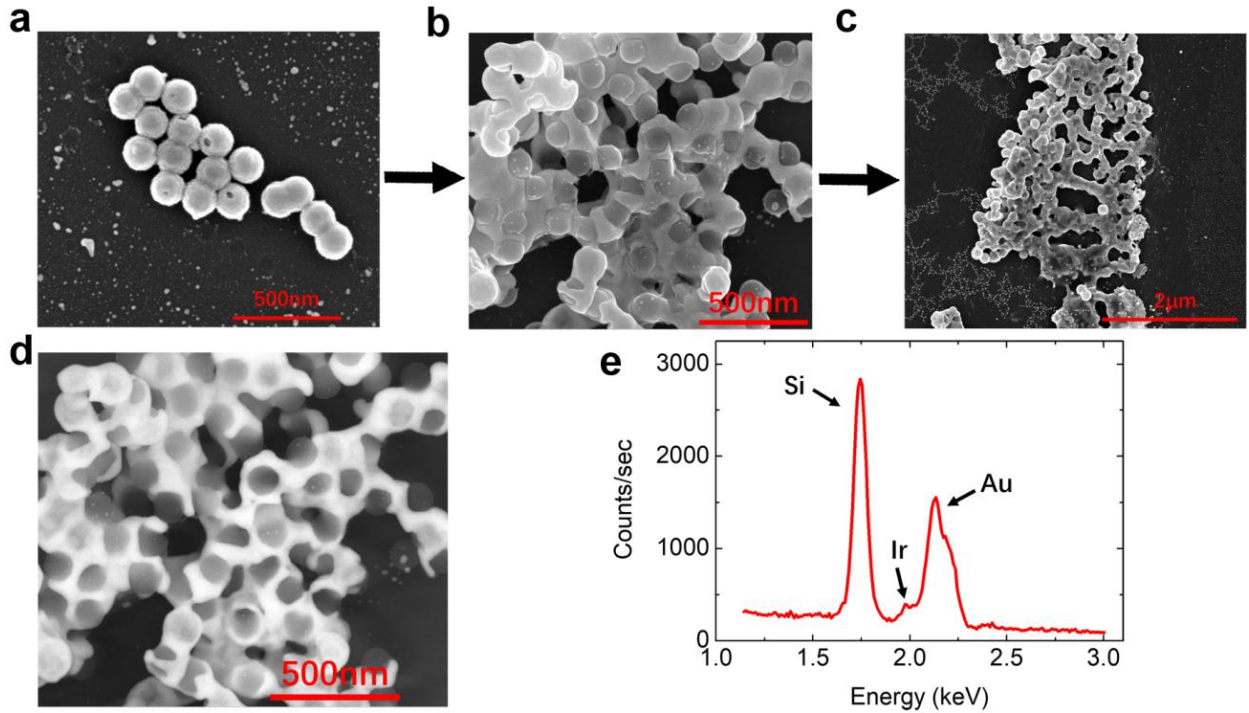

**Figure S2.** Scanning electron microscope (SEM) images of pre-deposited Au NPs at the bubble nucleation site in Case II. The bright dots in **(a)** are Au NPs. From **(a)** to **(b)**, we can observe the melting and merging of individual Au NP. In **(c)**, some Au clusters are formed. Scale bars are identified in figures. **(d)** Back-scattered SEM image of the same area in **(b)**. The bright region is Au and the dark dots are silica cores of the Au NPs. **(e)** Energy-dispersive X-ray (EDX) spectrum of the pre-deposited Au NPs at the bubble nucleation site in Case II. The corresponding element of each peak is identified.

**SI2. Surface bubble shrinkage study**

Figure S3 shows the bubble volume shrinkage plot. It is clear that the shrinkage is steady and linear, which agrees with the shrinkage behavior of gas bubbles. According to the theoretical model in Ref. [19], the volume shrinkage rate ($K_{shrink}$) of a gas bubble is:

$$K_{shrink} = \frac{6RTD\gamma}{P_\infty K_H} \tag{1}$$

where $R$ is the gas constant, $T$ is temperature, $D$ is laser beam diameter, $\gamma$ is surface tension of air/water interface, $P_\infty$ is ambient pressure and $K_H$ is the Henry coefficient of air in water. The volume shrinkage rate fitted in our case is about 420 μm$^3$/s, which is of the same order of magnitude as the calculated one from equation (1).

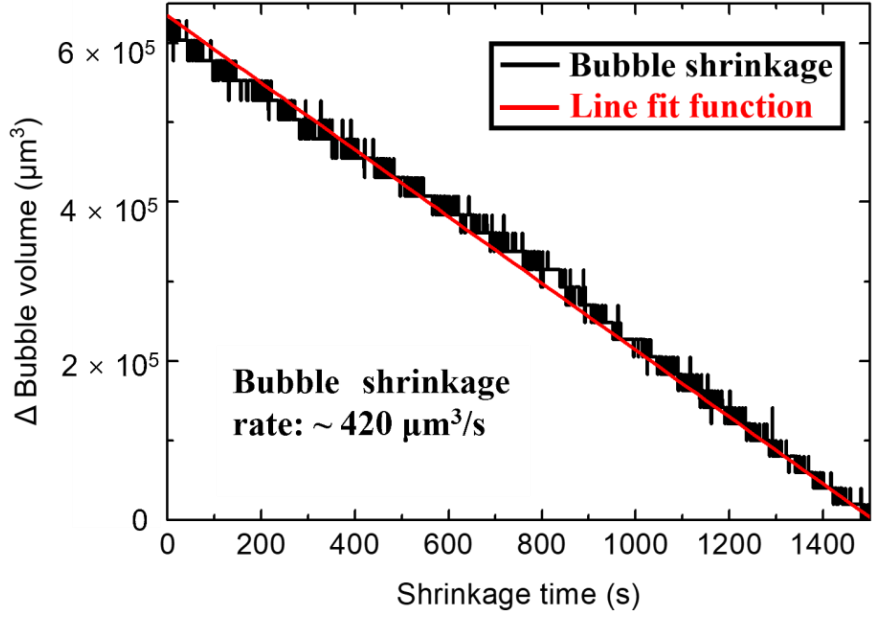

**Figure S3.** Bubble volume shrinkage (black) and the line fit (red) as a function of time. The bubble volume shrinkage rate is ~ 420 µm³/s.

## SI3. Finite element simulation details

We employ *COMSOL Multiphysics* to simulate the temperature profiles and the flow velocity of the surface bubbles in the volumetric heating and surface heating geometries. Both the flow effect and thermal conduction are included in our simulations. The models and boundary conditions used in our simulations are illustrated in Figure S4. In the volumetric heating simulation, as shown in Figures S4a and b, the heating intensity follows a Gaussian distribution as the laser intensity profile with a heat generation rate of:

$$Q_v = \eta_{abs}\alpha \frac{P_0}{2\pi\sigma^2} exp\left[-\left(\frac{(r-d)^2}{2\sigma^2} + \alpha(z-h)\right)\right] \quad (2)$$

where $\eta_{abs}$ ~ 0.2 is the optical absorption efficiency of Au NPs, which is determined by the ratio of the absorption quality factor and the extinction quality factor of the Au NP in DI water, as shown in ref. [28]. The optical attenuation factor of the NPs suspension $\alpha$ is ~ 262 m⁻¹, which is extracted from the absorbance spectrum in Figure 5b by the formulas:

$$Adsorbance(\lambda) = log^{\frac{1}{T(\lambda)}} \quad (3)$$

$$T(\lambda) = e^{-\alpha Z_0} \quad (4)$$

where $\lambda$ is the resonant laser wavelength of Au NPs in DI water, which is ~ 780 nm. $Adsorbance(\lambda)$ is the absorbance amplitude of the NP concentration of $2 \times 10^{15}$ particles/m³ in Figure 5b at the resonant wavelength. $T(\lambda)$ is the transmission of laser at the resonant wavelength. $Z_0$ is the length of laser path in NPs suspension, which is 1 cm. For our 10× objective lens, $\sigma = 11$ μm is the width of the Gaussian laser beam. $d$ represents the distance from the bubble central axis, and $h$ is the height of the bubble in the z-direction from the quartz surface at the distance $d$ from the center. In the surface heating geometry (as shown in Figures S4c and d), we use a thin layer of SiO$_2$ (10 μm-thick, 20 μm-width) sitting at the bottom of a surface bubble with a radius of 120 μm as the bottom heating source. The heat generation rate is:

$$Q_b = \frac{fP_0}{V} \tag{5}$$

where $P_0$ is the source laser power, $V$ is the volume of the thin film heater mimicking the deposited NPs, and $f$ is the portion of laser power which is used to heat the bubble. To determine the value of $f$, we fit our experimental phase II bubble volume growth rate (K) in Case II with the theoretical model described in Ref. [17]:

$$K = \frac{1}{3}\frac{RT}{M_g P_\infty}\frac{C_\infty}{C_s}\left|\frac{dC_s}{dT}\right|\frac{fP_0}{c_w\rho} \tag{6}$$

where $M_g$ is the molecular mass of air, $c_w$ is the specific heat capacity of water and $\rho$ is the density of water. The fitted value of $f$ is ~ 0.2%. In our simulations, an extremely fine mesh is used in both volumetric heating and surface heating geometries. The mesh structures of volumetric heating and surface heating geometries are shown in Figures S5 and S6 accordingly. Based on our convergence tests, as shown in Figure S7, the numbers of mesh elements used in the two geometries can provide sufficient accuracy.

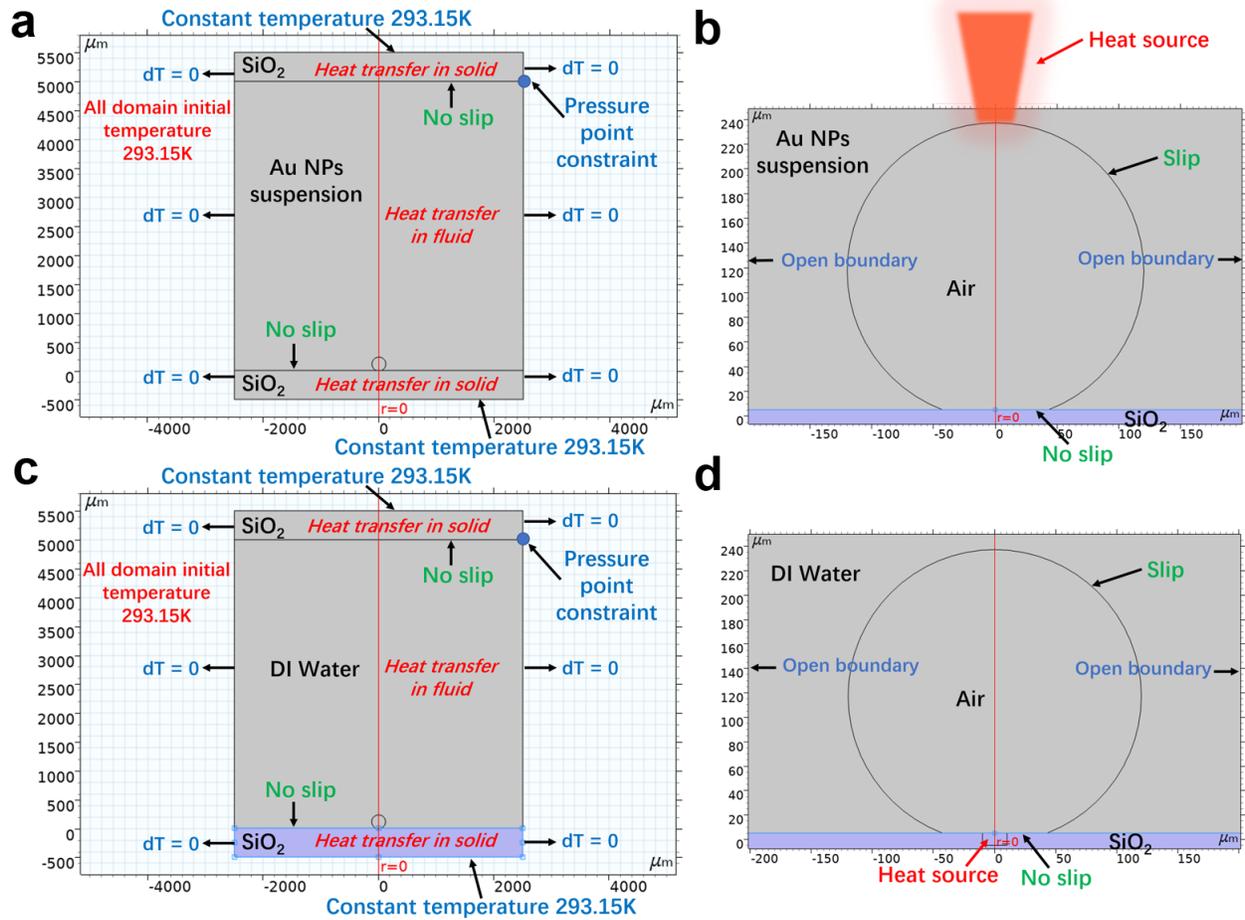

**Figure S4.** Schematic structures and boundary conditions for the simulations of the temperature profiles and the flow velocity profiles around the surface bubbles in the volumetric heating (**a, b**) and surface heating geometries (**c, d**). (**b**) and (**d**) zoom in the surface bubble regions of (**a**) and (**c**), respectively.

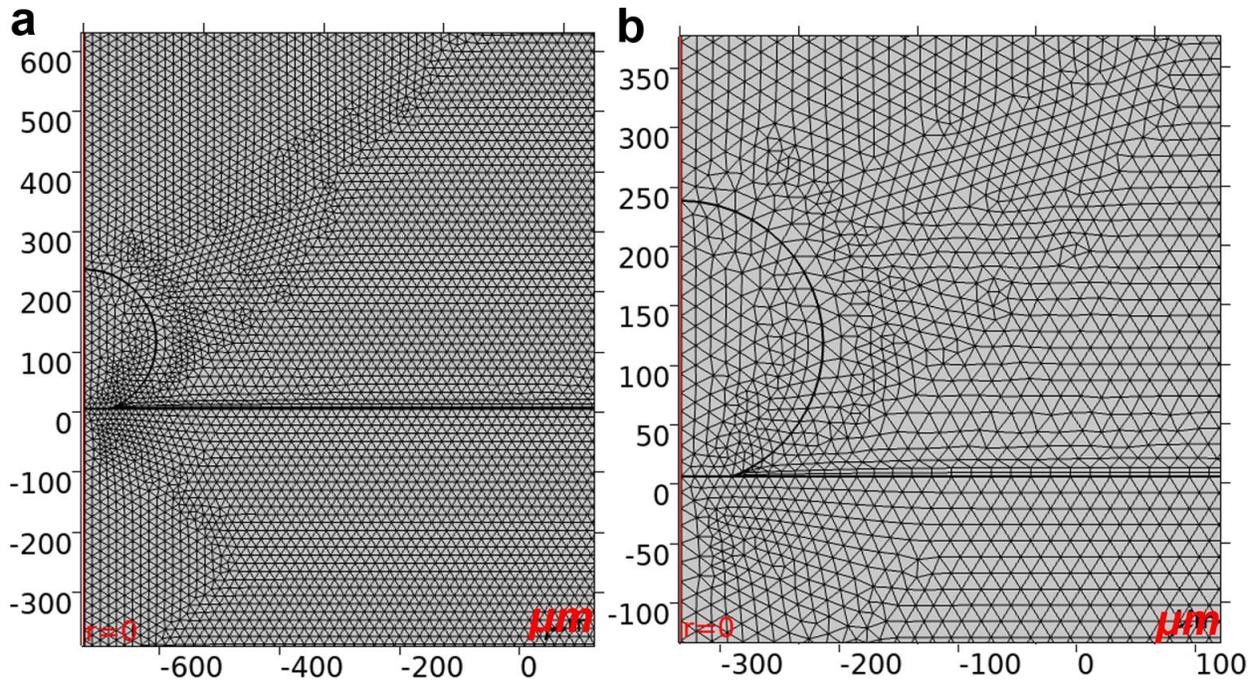

**Figure S5. (a)** The mesh structure employed in the simulation of volumetric heating geometry. The total number of mesh elements is 134812. **(b)** zooms in the surface bubble region of **(a)**.

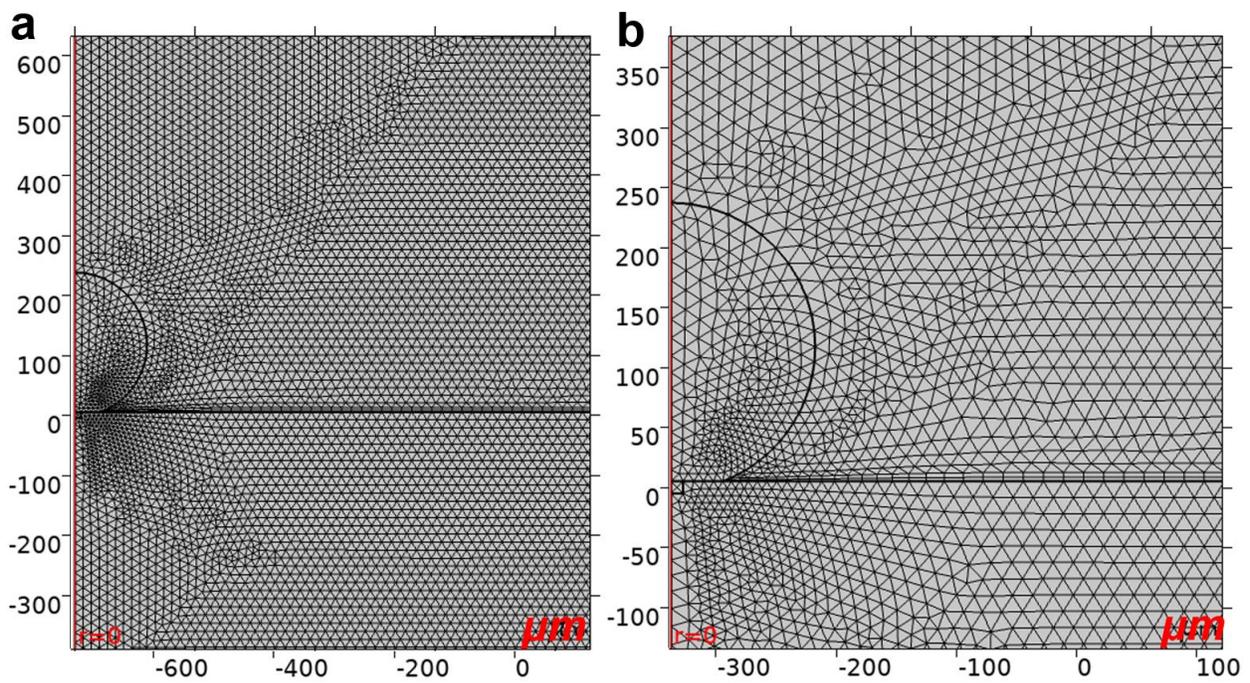

**Figure S6. (a)** The mesh structure employed in the simulation of surface heating geometry. The total number of mesh elements is 136602. **(b)** zooms in the surface bubble region of **(a)**.

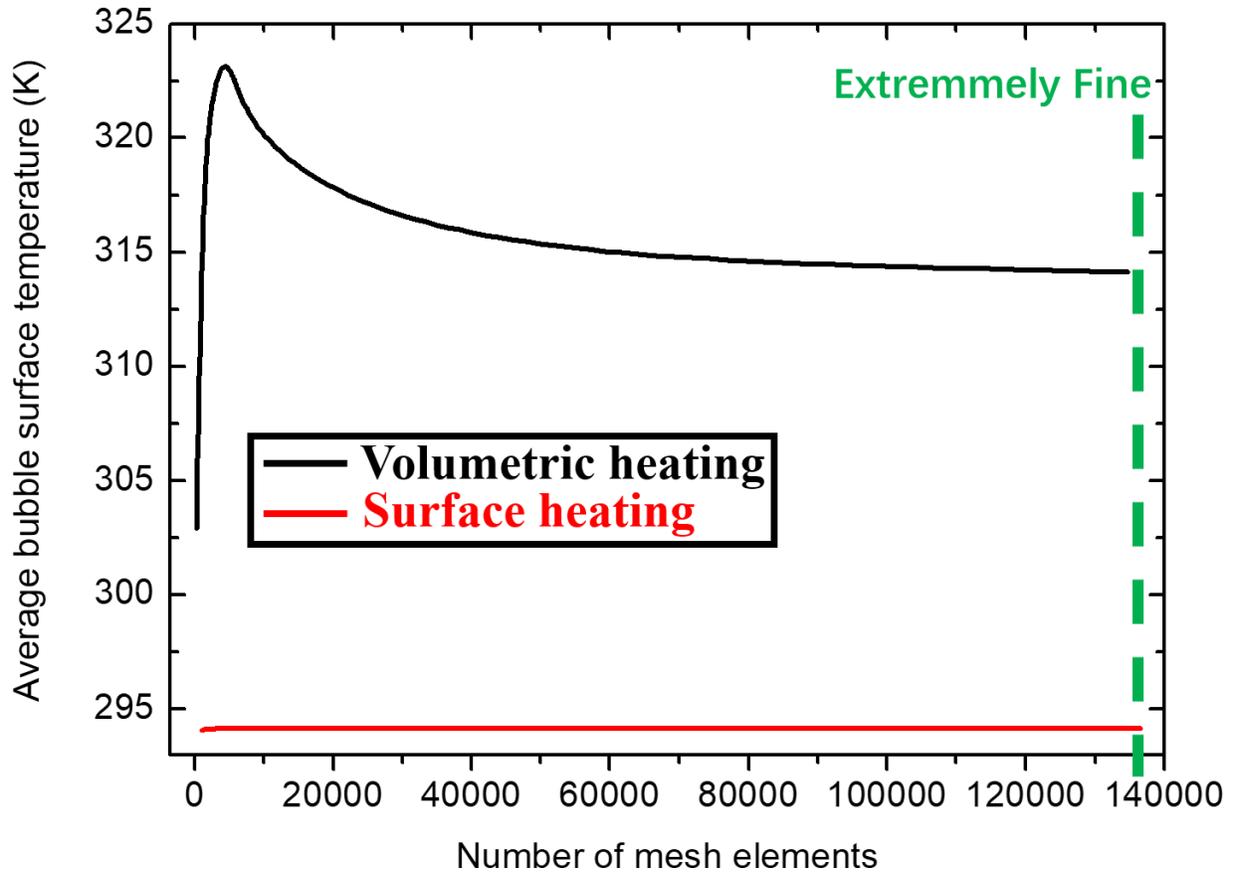

**Figure S7.** The convergence tests of the mesh structures used in the simulations of both volumetric heating and surface heating geometries. In the figure, the average bubble surface temperatures in the two heating geometries are plotted as a function of the number of mesh elements used. In our simulations, an extremely fine mesh structure is used in both heating geometries (as indicated in a green line).